\documentclass[aps,pra,twocolumn,notitlepage,superscriptaddress]{revtex4-1}

\usepackage{physics}
\usepackage{xcolor,graphicx}
\usepackage{comment}
\usepackage{siunitx}
\usepackage[normalem]{ulem}

\usepackage{amsthm,amssymb}
\usepackage{array}
\usepackage{hyperref}
\usepackage{mathtools}
\usepackage{bm,times,enumitem}
\usepackage{physics}
\usepackage{algorithm}
\usepackage{algc}
\usepackage{algcompatible}
\usepackage{algpseudocode}
\usepackage{soul}
\usepackage{dsfont}

\begin{document}

\title{Optimizing quantum-enhanced Bayesian multiparameter estimation of phase and noise in practical sensors}
\author{Federico Belliardo}
\affiliation{NEST, Scuola Normale Superiore and Istituto Nanoscienze-CNR, I-56126 Pisa, Italy}

\author{Valeria Cimini}
\affiliation{Dipartimento di Fisica, Sapienza Universit\`{a} di Roma, Piazzale Aldo Moro 5, I-00185 Roma, Italy}

\author{Emanuele Polino}
\affiliation{Dipartimento di Fisica, Sapienza Universit\`{a} di Roma, Piazzale Aldo Moro 5, I-00185 Roma, Italy}

\author{Francesco Hoch}
\affiliation{Dipartimento di Fisica, Sapienza Universit\`{a} di Roma, Piazzale Aldo Moro 5, I-00185 Roma, Italy}

\author{Bruno Piccirillo}
\affiliation{Department of Physics ``E. Pancini'', Universit\'a di Napoli ``Federico II'', Complesso Universitario MSA, via Cintia, 80126, Napoli}
\affiliation{INFN, Sez. di Napoli, Complesso Universitario di Monte Sant'Angelo, via Cinthia, 80126 Napoli, Italy}

\author{Nicol\`o Spagnolo}
\affiliation{Dipartimento di Fisica, Sapienza Universit\`{a} di Roma, Piazzale Aldo Moro 5, I-00185 Roma, Italy}

\author{Vittorio Giovannetti}
\email{vittorio.giovannetti@sns.it}
\affiliation{NEST, Scuola Normale Superiore and Istituto Nanoscienze-CNR, I-56126 Pisa, Italy}

\author{Fabio Sciarrino}
\email{fabio.sciarrino@uniroma1.it}
\affiliation{Dipartimento di Fisica, Sapienza Universit\`{a} di Roma, Piazzale Aldo Moro 5, I-00185 Roma, Italy}

\begin{abstract}
Achieving quantum-enhanced performances when measuring unknown quantities requires developing suitable methodologies for practical scenarios, that include noise and the availability of a limited amount of resources. Here, we report on the optimization of sub-standard quantum limit Bayesian multiparameter estimation in a scenario where a subset of the parameters describes unavoidable noise processes in an experimental photonic sensor. We explore how the optimization of the estimation changes depending on which parameters are either of interest or are treated as nuisance ones. Our results show that optimizing the multiparameter approach in noisy apparata represents a significant tool to fully exploit the potential of practical sensors operating beyond the standard quantum limit for broad resources range.

\end{abstract}

\maketitle
\section*{Introduction}

The goal of quantum metrology is to estimate a set of physical parameters, exploiting quantum resources to achieve improved performances beyond those achievable by classical methods. The use of quantum probes discloses the capability to reach the Heisenberg limit (HL), gaining a quadratic scaling advantage over the standard quantum limit (SQL) corresponding to the use of $N$ independent probes \cite{Giovannetti1330,PhysRevLett.96.010401,barbieri2022optical,Giovannetti,avsreview2020}. Often in a real scenario, even if the interest relies on a single parameter, the process is unavoidably affected by the presence of unknown noises. For these reasons, it is usually more effective to treat these estimations using a multiparameter approach   \cite{liu2019quantum,suzuki2020quantum,suzuki2020nuisance,multiparameter_QM,albarelli2020perspective}. Despite their importance, experimental demonstrations of quantum enhanced estimation in the multiparameter case are still few and limited to modest amounts of coherent quantum resources
\cite{polino2019experimental,liu2021distributed,hong2021quantum,valeri2023experimental,cimini2022deep,avsreview2020}. The following two extremal scenarios are prototypical for multiparameter metrology. In the first case, all the unknown parameters are treated on the same level, and thus one needs to optimize the overall amount of information extracted. Here, the adoption of quantum probes can provide improved performances with respect to strategies where each parameter is estimated separately \cite{humphreys2013quantum,PhysRevLett.128.040504,belliardo2021incompatibility,Yue}. In the second extremal case, only one parameter is considered to be of interest, but the dynamics of the metrological evolution intrinsically involves other nuisance parameters, of which an approximate knowledge is although necessary to retrieve a good estimator for the desired one. For instance, we have to deal with this scenario when different sources of noise affect the evolution: phase and visibility \cite{Roccia:18,roccia2017entangling,cimini2019quantum,cimini2019adaptive}, phase and phase diffusion \cite{Vidrighin,PhysRevLett.106.153603}, magnetic field and decoherence \cite{PhysRevA.84.012103} for example. The optimal strategy in this case is very different from the optimal one in the former scenario, since now the interest is to maximize the information extracted on one parameter at the expense of all the others \cite{goldberg2020multiphase}. In the general case, intermediate configurations between these two extremal scenarios can be defined, corresponding to different choices of the cost function. For example, a couple of parameters could be considered of interest while the others are treated as nuisance. For each specific scenario, different strategies may thus turn out to be optimal. In general, the importance of the different parameters can be weighted arbitrarily.

Another crucial aspect of quantum metrology in a practical scenario regards the availability of a finite amount of resources $N$ in the estimation process. The standard approach is based on a theoretical framework dedicated to defining bounds and strategies in the asymptotic limit of large $N$. However, when only a finite amount of resources is available, any estimation strategy needs to be tailored to optimize the convergence for low values of $N$ \cite{rubio2020quantum,rubio2019limited,rubio2020bayesian}. A powerful tool here is represented by adaptive protocols, which enable faster convergences to the ultimate limits \cite{wiseman1995adaptive,berry2000optimal}. These have been implemented both through online \cite{berni2015ab,Cimini:19} and offline \cite{hentschel2011efficient,lovett2013differential} approaches also resorting to the use of different machine learning algorithms \cite{hentschel2009adaptive,piccoloLume,rambhatla2020adaptive,cimini2023deep}. These techniques demonstrated two relevant characteristics, namely fast convergence to the ultimate bounds and performances independent of the particular value of the parameter of interest. Different experimental applications of adaptive techniques have been reported, first in single-parameter estimation problems \cite{armen2002adaptive,higgins2007entanglement,daryanoosh2018experimental,wheatley2010adaptive,piccoloLume,rambhatla2020adaptive} and then in a multiparameter setting \cite{Valeri2020,valeri2023experimental}. In this scenario, the versatility of the multiparameter approach allows to choose the optimal allocation of resources, depending on which are the parameters of interest and which are the ones associated to noise processes, treated instead as nuisances. Furthermore, the application of adaptive strategies in the quantum metrology context is beneficial also when the quantum probe state is in a continuous variable (CV) Gaussian state. In particular, squeezed light has emerged as a valuable resource as demonstrated by its use in various domains, from gravitational wave detection\cite{tse2019quantum} to recent research on distributed quantum sensing, where CV states enable enhanced precision and sensitivity in multiparameter estimation tasks \cite{guo2020distributed,xia2020demonstration}.
However, also in the CV framework, real-world measurement are often dynamic and subject to various sources of noise and fluctuation. Adaptive strategies, such as adaptive Bayesian estimation, allow systems to dynamically respond to changing conditions, optimizing measurement protocols in real-time ensuring enhanced measurement performances. 

In this work, we investigate a multiparameter estimation scenario, where the parameters of interest are a physical rotation angle~\cite{goldberg2018quantum,goldberg2021rotation} together with the noise values involved in the interferometric measurements. To this end, we employ orbital angular momentum (OAM) of single photons, carrying tunable OAM values up to $50$, able to show N00N-like sensitivities for rotation estimations \cite{dambrosio_gear2013,cimini2023experimental,Fickler13642,barnett2006resolution,jha2011supersensitive,PhysRevLett.127.263601}. Importantly, we extend the single parameter study \cite{cimini2023experimental} to a multiparameter approach within a Bayesian framework \cite{helstrom1976quantum,box2011bayesian,d2022experimental,rubio2020bayesian} for all the aforementioned scenarios by employing an adaptive strategy ensuring the optimal allocation of resources \cite{granade2012robust}. Such an approach allows us to extend the multiparameter estimation problems to the regime of $O(30,000)$ number of quantum-like resources, experimentally demonstrating sub-SQL precision in the estimation of the rotation angle for wide resources ranges even with nuisance parameters. In order to quantify the quality of the reached performances, we define non-tight Bayesian bounds on the estimations in each considered scenario.
This work is applicable to all those cases where the visibilities must be estimated on the fly, for example because they fluctuate from one rotation estimation to the other. However, also if a pre-calibration of the visibilities before the rotation estimation is possible, our approach will be optimal with respect to the figures of merit and resources considered, if the resources consumed in the pre-calibration are also counted. Indeed, if a pre-calibration would be more efficient, our protocol would choose it, as a first stage of the estimation.

\section*{Precision bounds for the multiparameter estimation}
The goal of multiparameter quantum metrology is to identify regimes where the estimation precision outperforms the one achievable by classical probes. A crucial aspect is to keep such enhanced performances as the employed resources increase. It becomes key to develop a platform able to investigate such a regime with quantum scaling of the precision. Here, we study the simultaneous estimation, in the large resource regime, of a rotation angle $\theta\in [0,\pi)$ and of a collection of parameters (the fringe visibilities $V_{s_1}$, $\cdots$, $V_{s_4}$ defined below) that affect the efficiency of the detection process~\cite{dambrosio_gear2013}. More specifically, in our scheme, before each measurement step, we select a control parameter $s$ out of a set of four possible values $\{s_1=1,s_2=2,s_3=11,s_4=51\}$, each corresponding to a device which can be switched on at will and produces the quantum-like resource. In the ideal (noiseless) scenario such choice is meant to force the interferometer to produce a single-photon, N00N-like output state analogous to those employed in \cite{higgins2007entanglement} to achieve quantum limited precision, i.e. the vector $|\psi_s(\theta)\rangle:=\frac{1}{\sqrt{2}} (\ket{0}+e^{-2is\theta}\ket{1})$, with $|0\rangle$, $|1\rangle$ being orthogonal circular polarization states. Unfortunately, the selection of high values of $s$ also has the indirect effect to add noise into the model which ultimately deteriorates the associated visibility $V_s\in[0,1]$ of the measurements we perform on $|\psi_s(\theta)\rangle$ to recover $\theta$. Our scheme relays on two different types of detections, the first corresponding to the projection of $|\psi_s(\theta)\rangle$ on the basis $\{ (\ket{0}\pm \ket{1})/\sqrt{2}\}$, while the second uses $\{ (\ket{0}\pm i \ket{1})/\sqrt{2}\}$ as reference basis. Our interest lies in determining how the estimation precision changes depending on the different perspective from which the multiparameter problem is addressed. To this end, we apply the Bayesian algorithm in Ref.~\cite{granade2012robust}. 
This procedure identifies the most effective adaptive strategy depending on the different roles assigned to each parameter ($\theta$, $V_{s_1}$, $V_{s_2}$, $V_{s_3}$, and $V_{s_4}$), whether they are of interest or are treated as nuisance parameters. In this protocol, the Bayesian posterior probability distribution for all the parameters is represented by a particle filter. 
%
%
For more details see Appendix B and~\cite{granade2012robust}. From the information accumulated in the reconstructed posterior distribution, at each step, a greedy strategy selects the experimental settings that minimize the future expected estimator variance \cite{granade2012robust}. This is computed through a brute force procedure that consists in simulating the evolution of the posterior distribution for all the possible measurements settings, and picking the one with the smaller expected variance. We optimize both the value $s$, that determines the successive probe state and the basis of the polarization measurement we perform on the output state $|\psi_s(\theta)\rangle$.
The information flow in the Bayesian algorithm is represented in Fig.~\ref{fig:setup}, here the measurement outcomes are used to update the posterior by the computer unit that also calculates the next optimal measurement. The parameters tuned by the optimization algorithm are the used quantum resource and the appropriate polarization measurement basis. To quantify the efficiency of the estimation, we define a suitable figure of merit and derive bounds limiting its minimal achievable value for the investigated metrological task.
Focusing on a generic experimental run ${{\mathbf r}_N}$ composed by a series of individual measurements, where, given $i\in \{1,\cdots, 4\}$, the control value $s_i$ is used $\nu_i$ times, we define the total number of resources $N =\sum_{i=1}^4 \nu_i s_i$ ~\cite{PhysRevLett.96.010401}.
Hence, indicating  with $\hat{\theta}^{({\mathbf r}_N)}, \hat{V}^{({\mathbf r}_N)}_{s_1}, \cdots, \hat{V}^{({\mathbf r}_N)}_{s_4}$ the estimated values of ${\theta}, {V}_{s_1}, \cdots, {V}_{s_4}$, we get from the Bayesian procedure, we gauge the associated experimental error via the quantity $\Delta^{2}_{{\mathbf r}_N,G}(\theta):=G_{1,1}|\hat{\theta}^{({\mathbf r}_N)}-\theta|^2 + \sum_{i=1}^4 G_{i+1\, , i+1} |\hat{V}^{({\mathbf r}_N)}_{s_i}-V_{s_i}|^2$,
where $G$ is a weight matrix that controls which parameters are to be treated as nuisance and which are the parameters of interest~\cite{goldberg2020multiphase}.
We repeat the whole estimation for a collection $\theta_1, \theta_2,\cdots,\theta_J$ of $J$ different values of $\theta$ uniformly distributed in $[0, \pi)$, and for each $\theta_j$, we run the whole procedure $M>200$ times. 
Indeed, in this protocol, we can not fix the total number $N$ of resources, because the stochastic nature of the measurement outcomes propagates to the online choice of the next q-plate and then to $N$, which is, therefore, a stochastic variable whose exact value is only known at runtime. However, we can repeat $M \gg 1$ times the simulation and collet the precision after each used photon in the set of tuple $\lbrace (\Delta_{{\boldsymbol{r}_n}, G}^2 (\theta_j), n) \rbrace$, where $n$ is the total resource number used up to that point to get to precision $\Delta_{{\boldsymbol{r}_n}, G}^2 (\theta_j)$. We will then look at all the points with total resources falling in the interval $[n, n+\Delta n]$ with $\Delta n$ small, and the median error of this cluster is the error we associate with $n$. We choose to cluster only the point having $n>100$, with the chosen interval being $\Delta n = 50$. 
This finally leads us to the identification of the error figure of merit:
\begin{equation}
\mathcal{M}^2_G := \text{Median} \Big[ \frac{1}{J}
	\sum_{j=1}^J \Delta^{2}_{{\mathbf r}_N,G}(\theta_j) \Big] \; ,
	\label{eq:medianMerit}
\end{equation}
with the median taken on the $M$ executions. This is the median square error over the different estimates, each obtained, as usual in quantum metrology, as the mean of the posterior distribution. The choice to compute the median instead of the average is due to the fact, that the former is much less sensitive to outliers which are always present in  such estimation protocols (see Appendix A). 
It is possible to impose a theoretical lower limit on~(\ref{eq:medianMerit}) which can be used to evaluate the quality of our experiment. Specifically, we can write:

\begin{equation}
	\mathcal{M}^2_G \ge {\widetilde{C_G}}/{N} \; ,
	\label{eq:boundMSE}
\end{equation}
where the constant $\widetilde{C_G}$  is defined and computed in Appendix A via non-linear programming on the Cramér-Rao bound and connecting the mean square error to the median square error. This bound is legit if we assume the local asymptotic normality (LAN) of the estimators. 

\section*{Experimental apparatus}
\begin{figure}[!t]
	\includegraphics[width=0.99\columnwidth]{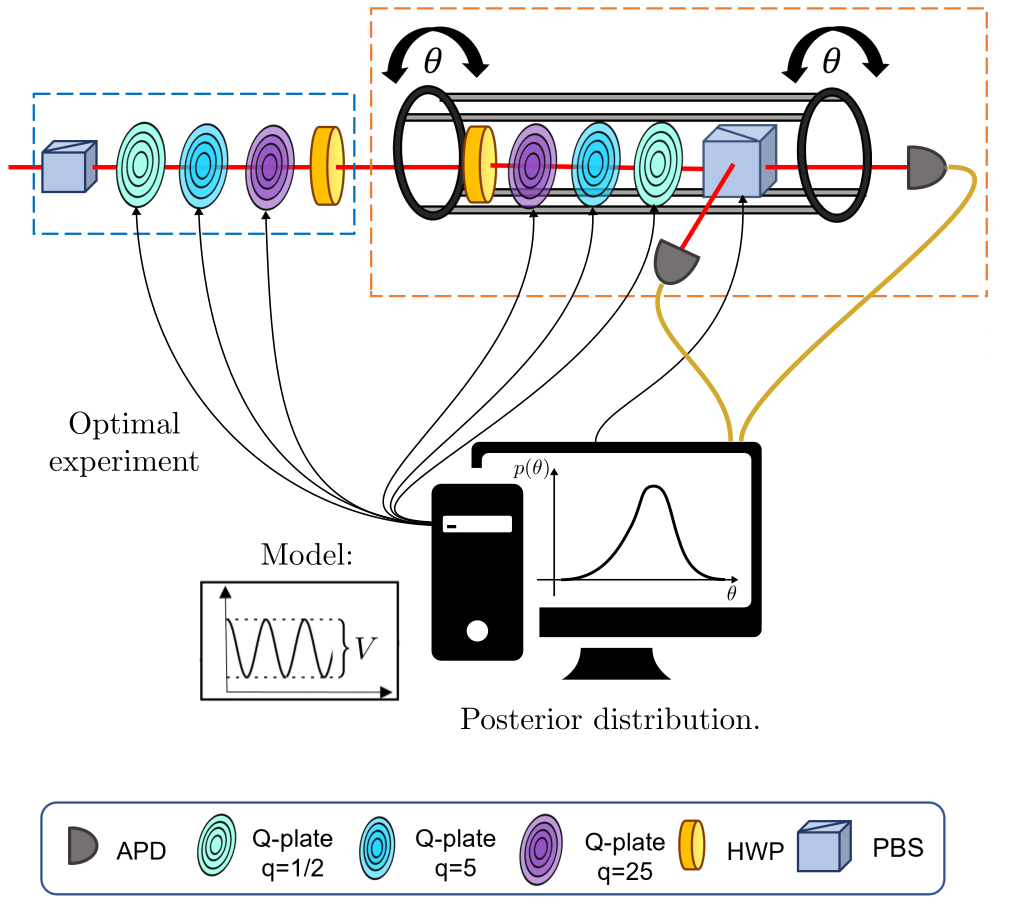}
	\caption{Sketch of the experimental setup. Single photons are sent through the apparatus consisting of a generation stage (blue rectangle) and a measurement stage (orange rectangle). The first is composed of a polarizing beam splitter (PBS) and three q-plates with different topological charges $q = 1/2; 5; 25$, respectively, followed by a half-waveplate (HWP). The measurement stage is composed of the same elements of the preparation mounted in reverse order in a compact and motorized cage which can be rotated by an angle $\theta$. After the final PBS, the photons are measured through avalanche photodiodes (APDs). The measurement HWP allows to set the polarization measurement basis. The outcomes of the APDs are used by a computer to update the posterior distribution and optimize, for the next measurement, the tunable parameters consisting of the activated q-plate and the polarization basis.}
	\label{fig:setup}
\end{figure}

In order to investigate a multiparameter metrological problem, we employ a state-of-the-art experimental apparatus, that allows to generate single-photon high-OAM value states showing quantum-like sensitivities for the estimation of a rotation angle. Importantly, we extend the result from the single parameter case \cite{cimini2023experimental} to a multiparameter one, exploring unprecedented regimes for multiparameter quantum metrology. The experimental setup is composed of a series of q-plates \cite{marrucci-2006spin-to-orbital,cimini2023experimental} with an increasing topological charge $q$ arranged in a cascade configuration as reported in Fig.~\ref{fig:setup}. We test the protocol through single-photon states generated via a spontaneous parametric down-conversion (SPDC) source in a Sagnac configuration. Photon pairs are emitted at $808$ nm pumping a periodically poled titanyl phosphate (ppKTP) crystal with a continuous laser with a wavelength equal to $404$ nm. Once generated, one photon is sent through the apparatus while the other acts like a trigger allowing to work with coincidences events when the two photons of the pairs are detected by avalanche photodiodes within a time window of few nanoseconds. Starting from single photons prepared in the horizontal polarization state, we can generate a N00N-like state in the OAM degree of freedom of the form: $\frac{1}{\sqrt{2}}(\ket{0}\ket{m}+\ket{1}\ket{-m})$, where $\ket{0,1}$ refers to the circular polarization, while $\ket{\pm m}$ with $m=2q$ represents the OAM state and depends on the activated q-plate. The prepared state is then sent to a measurement stage composed of the same set of q-plates in reverse order, allowing to reconvert the OAM state into a polarization state. Such measurement stage can be rotated by an angle $\theta\in[0, \pi)$ by means of a fully motorized rotation cage, as realized in Ref. \cite{cimini2023experimental}. In this way, the photon state passing through the full setup becomes the vector $|\psi_s(\theta)\rangle$ defined earlier, where $s=2q+1$ is the total angular momentum of the photon. By appropriately choosing the active q-plate, determining $m$, the frequency of the oscillation interference fringes can be tuned. Finally, through a half-waveplate after the q-plate, it is possible to select also the measurement polarization basis. 
Given the employed devices, we have access to states with $s = 1,2,11,51$. The first is obtained when no q-plate is activated, while the others are achieved activating in turn one and only one of the three mounted q-plates in the preparation stage, and the corresponding one in the receiver. These states generate oscillation patterns, retrieved from measurements in polarization basis and characterized by visibilities $V_s$ which depend upon the selected $s$, this is schematically represented by the insert ``Model'' in Fig.~\ref{fig:setup}.

\section*{Experimental results}

Here, we investigate the two scenarios mentioned in the introduction: in the first one the interest is focused only on the rotation angle while the visibilities involved in the process are only considered as unavoidable noises. In order to achieve a reliable estimation of the rotation, it is indeed useful to keep track of all the visibilities values which are therefore treated in such a scenario as nuisance parameters. In the second scenario inspected, we consider also the visibilities as interesting parameters. This happens for instance if the user is interested in the complete characterization of an interferometer, and therefore needs an estimation of the noise levels too. In this context, depending on the selected configuration, the optimal multiparameter estimation protocol can be found by assigning positive weights to both the rotation angle and the visibilities. Whatever the scenario, the optimization is performed in a Bayesian framework that follows the strategy developed in \cite{granade2012robust}. Our procedure is however very different when it comes to counting the resources. A single experiment consumes a single photon, however, we do not count this as the consumption of a single resource but as the use of a number of resources equal to the total angular momentum of the photon in that experiment. This is done to set a correspondence between the photon states and the entangled N00N states with sizes equal to the available total angular momentum of the photons. Our N00N-like states can achieve quantum-limited precision, similarly as multipass protocols do \cite{higgins2007entanglement}.
It is worth noting that, similar to multipass protocols, which quantify the resources invested in estimation based on the number of interactions between the probe and the sample, it is reasonable to consider the total angular momentum as a valuable resource in the estimation protocol. There are several reasons for this choice. Generating and measuring higher-order orbital angular momentum requires more complex devices such as q-plates with higher topological charges or spatial light modulators with larger effective areas. Moreover, the propagation of these states requires also to address their divergence.

\begin{figure*}[!htb]
	\includegraphics[width=\textwidth]{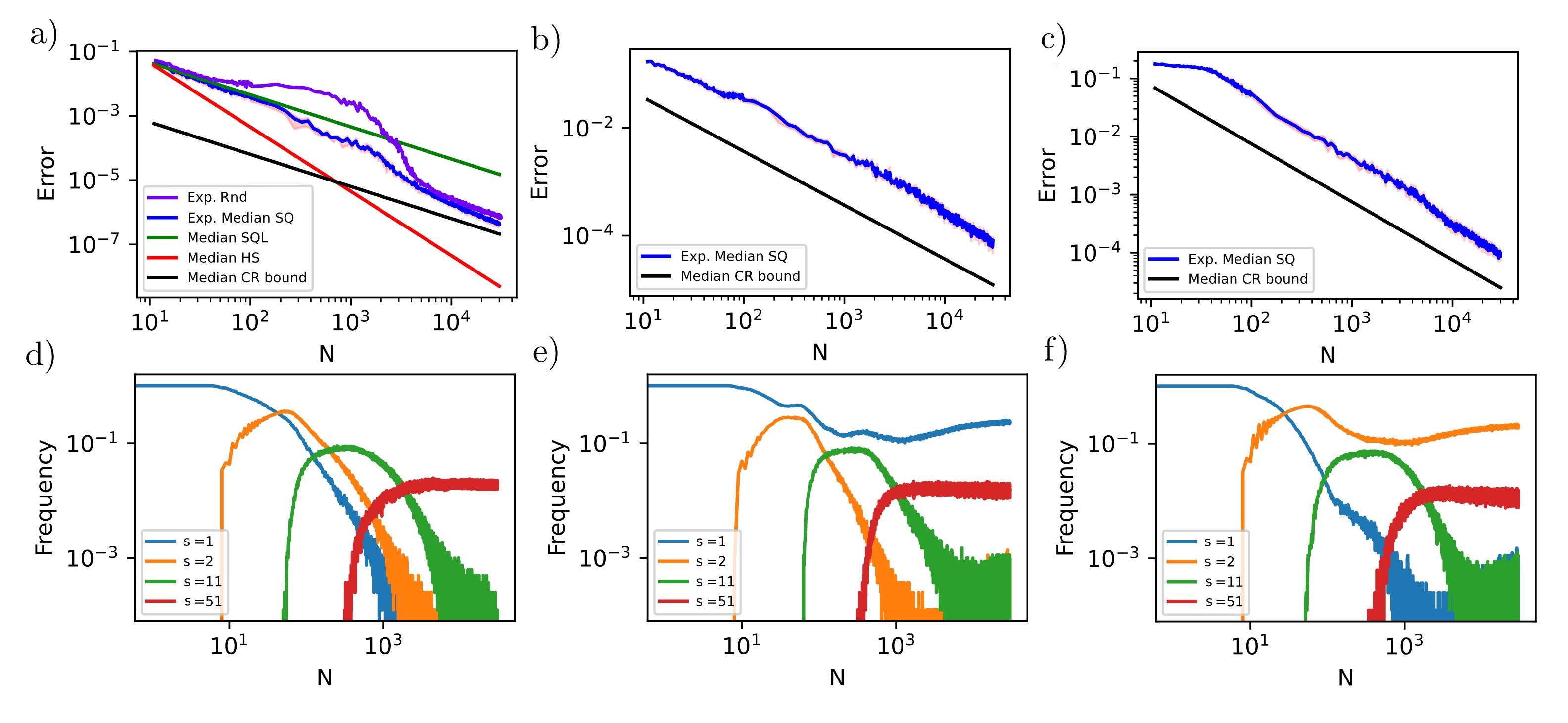}
	\caption{a) Median square error of Eq.\eqref{eq:medianMerit} for the phase only (blue line), with all the other parameters treated as nuisances. The green line is the SQL$=1/N$, and the red one is the HL$=\pi^2/N^2$, both converted to bound on the median square error. The violet line is the precision for non-adaptive measurements. In all the plots the black line is the bound of Eq.~\eqref{eq:boundMSE} computed for the appropriate weight matrix $G$. b)-c) Median square error for the phase and one of the visibilities, respectively $V_{s_1}$, $V_{s_2}$. For each plot, all the other three visibilities are treated as nuisance parameters. In light red we plot the 99\% confidence interval for the experimentally evaluated median error. d) - e) - f) Frequency of use of each q-plates as a function of the number of resources, computed for a batch of $8 \cdot 10^3$ experiments with the measurement optimization described in the main text.}
	\label{fig:phase}
\end{figure*}

We start measuring the angle $\theta$ while treating the visibilities as nuisance. To this end, we set the weight matrix $G$ in Eq.~\eqref{eq:medianMerit} to have $G_{11} = 1$ as the only non-null entry. The online execution of the Bayesian algorithm is only simulated. This means we first run a stage where we perform a large number of measurements for all the possible q-plates, polarization basis, and rotation angles $\theta_j$, and then we queue the outcomes. In the second stage, in running the Bayesian analysis offline, we evoke the needed measurement at each step from the appropriate queue. We collect such outcomes for $J=8$ different rotation angles in~$[0, \pi)$ (reported in the Table of Appendix \ref{tab:visibilities}) , which is the periodicity interval of the system, reporting in panel a) - b) and c) of Fig.~\ref{fig:phase} the experimental median squared error for some of the different investigated scenarios. 
The optimized median square error when the rotation angle is the only parameter of interest is reported in panel a) of Fig.~\ref{fig:phase} where the experimental data are compared with the Cramér-Rao bound on the median of Eq.~\eqref{eq:boundMSE}, the SQL and the HL ~\cite{PhysRevLett.124.030501} (both corrected to refer to the median error). Along with the precision for the optimized Bayesian strategy we report, as a comparison, also the experimental precision for a non-optimized strategy (violet line in Fig.~\ref{fig:phase}a)) in which the activated q-plate and polarization measurement basis are randomly chosen. While it shows for large resources an advantage with respect to the SQL, it is in the transient much worse than the optimized strategy.

In the Bayesian estimation the prior distributions on the visibilities is uniform in $[0, 1]$ and the prior on the angle is likewise uniform in $[0, \pi)$. With the goal of minimizing the posterior variance on the estimation of the rotation angle, the adaptive algorithm selects at each step the most appropriate quantum-like (q-plates) resource among the available ones. The resources that the Bayesian algorithm selects for the phase estimation are reported in panel d) - e) and f) of Fig.~\ref{fig:phase}. For $N>10^4$ only the q-plate with larger topological charge $q$ is used, while for $N$ very small only $s=1$ is selected. In between, we have a transient regime where multiple q-plates are used together. This is the region that benefits the most from the optimization. 

The experimental results show that the obtained error approaches the computed bound, which proves to be a valid reference even if non-tight. Notably, for the evaluation of the phase, even if the visibility values are completely unknown, the implemented multiparameter protocol shows an enhanced estimation precision compared to the SQL for a large resources range, previously unexplored by multiparameter estimation experiments. Importantly, differently from \cite{cimini2023experimental} (see Appendix C for more details) where the measurement strategy has been precalibrated according to the visibility values, here we show that it is still possible to obtain a large region showing sub-SQL performances in the medium-large resource range ($1<N<10^4$), treating the $4$ visibilities as nuisance parameters. 

\begin{figure*}[!htb]
	\includegraphics[width=\textwidth]{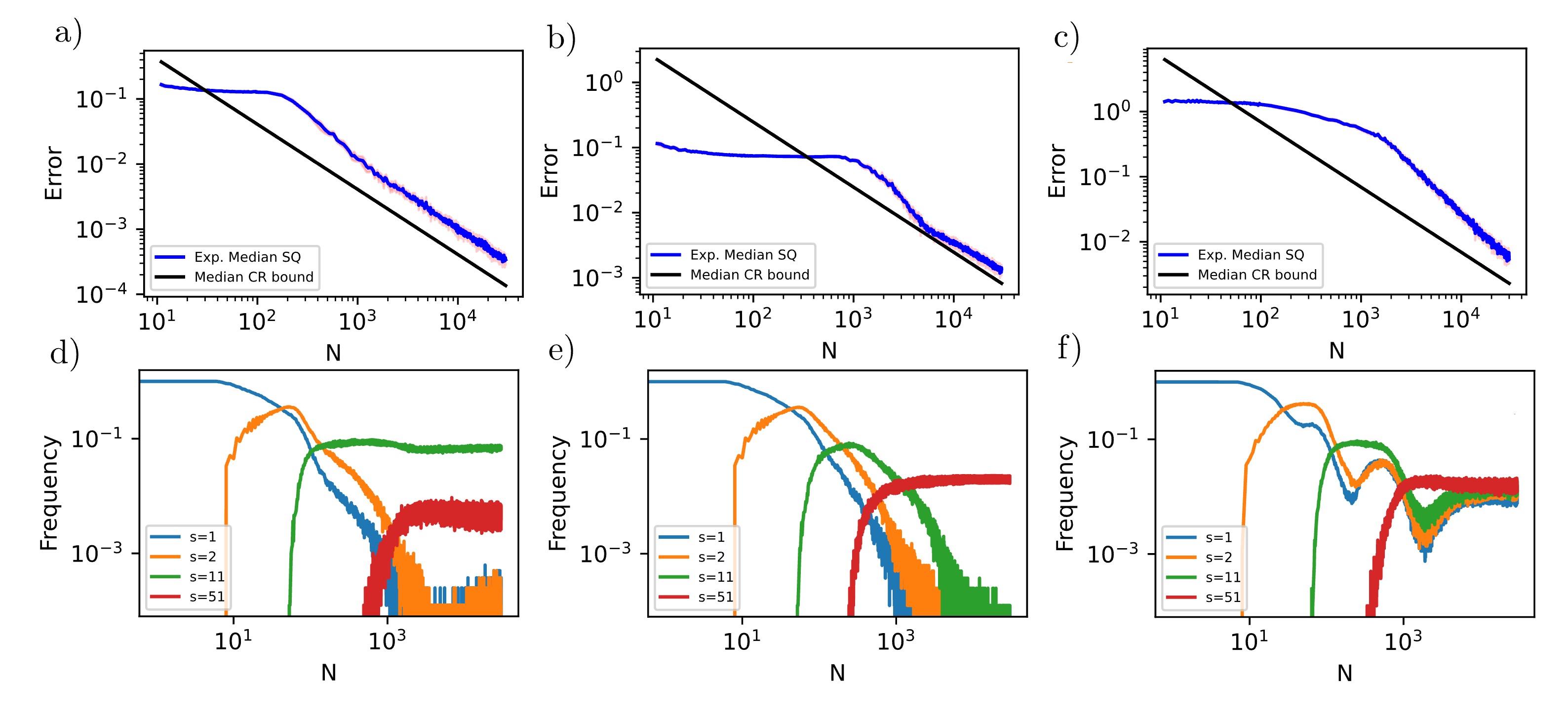}
	\caption{  a)-c) Median square error for the phase and one of the visibilities, respectively $V_{s_3}$ and $V_{s_4}$ (blue lines). For each plot, all the other three visibilities are instead treated as nuisance parameters. c) Median square error on the joint estimation of the phase and all four visibilities. In light red, we plot the 99\% confidence interval for the experimentally evaluated median error. d)-f) The frequency of use of each q-plates as a function of the number of resources, computed for a batch of $10^3$ experiments with the measurement optimization described in the main text. d)-e) refer to the joint estimation of the rotation angle and one visibility, respectively $V_{s_3}$ and $V_{s_4}$. Plot f) refers to the joint estimation of the rotation angle and all the visibilities. }
	\label{fig:phase_app}
\end{figure*}

We now consider the scenario where the visibilities are not treated anymore as nuisances but count in the evaluation of the error. This happens for instance if the user is interested in the full characterization of the apparatus and its configuration, therefore needing an estimation of the noise levels as well as of the rotation angle. It is interesting to see how the optimization of the available resources changes in these new configurations. In particular, we shall focus on the scenario where one is interested in the estimation of the angle and one of the visibilities (the $i_0$-th one), while the three remaining ones are still treated as nuisance parameters. Under this assumption, the parameters of interest are collected in the pair $(\theta, V_{s_{i_0}})$ and the median error of Eq.~(\ref{eq:medianMerit}) is computed with the weight matrix having $G_{1,1}=G_{i_0,i_0}=1$ as the only non-zero elements. As shown in Fig.~\ref{fig:phase}, under these circumstances the Bayesian protocol, although switching to higher dimensional OAM states to decrease the error on the rotation angle, continues to use also the q-plate related to the visibility chosen: this is necessary to obtain a good precision on the joint error. The results of the estimation of each of the four possible couples are reported in Fig.~\ref{fig:phase}b, c, d and e. In particular, the plateaus in Fig.~\ref{fig:phase}d and Fig.~\ref{fig:phase}e highlighted in green appear since, for few resources, the q-plates corresponding to $s=11$ and $s=51$ are not significantly used, therefore, the estimator of the corresponding visibilities remains the mean value of the uniform prior distribution in $[0, 1]$, i.e. $0.5$, while the error on the phase decreases, thereby reaching a plateau in the error determined by the value of the visibility $V_{i_0}$ itself. This changes only when the algorithm starts to use the high charge q-plates, and the error finally decreases.
In Fig.~\ref{fig:phase_app} other scenarios where the visibilities are not treated as nuisance parameters but become themselves parameters of study are reported. It is interesting to see how the optimization of the available resources changes in these new configurations. When we are interested in the estimation of the rotation angle and one of the visibilities, the Bayesian protocol, although switching to higher dimensional OAM states to decrease the error on the rotation angle, will continue to use the q-plate related to the visibility chosen (see Fig.~\ref{fig:phase_app} panel d)-e)).
In particular, the initial plateau appearing in Fig.~\ref{fig:phase_app} a)-b) are due to the fact that for few resources the q-plates corresponding to $s=11$ and $s=51$ are not significantly used and therefore the estimator of the corresponding visibilities remains the mean value of the uniform distribution in $[0, 1]$ (the prior), i.e. $0.5$, until when the algorithm starts to use the high charge q-plates, and the error finally decreases.
Lastly, we try to estimate all the five parameters $(\theta, V_1, V_2, V_3, V_4)$, that means setting all the diagonal elements of the matrix $G$ in equal to $1$, obtaining the results reported in panel c) and f) in Fig.~\ref{fig:phase_app}.

As a final observation, we notice that in the case where one focuses only on the phase $\theta$ there is still a pretty large region in which the slope of the optimized estimation error scales with sub-SQL performances (specifically looking at Fig.~\ref{fig:phase}a, this happens for values of $N$ between 2000 and 5000). The region where sub-SQL scaling (in $N$) can be observed is defined by the maximum amount of the topological charge $q$ employed in the apparatus. This region can be thus extended by increasing the maximum value of $q$. Such behaviour on the contrary is damped (but still present) when considering the estimation of the pairs $(\theta, V_{s_{i_0}})$. The reason for this is that the visibility is an inherently classical parameter and cannot benefit from the high angular momenta being our quantum resources. 

\section*{Discussions and conclusions}

In summary, quantum sensing promises to be one of the first quantum technologies exploited to enhance tasks with respect to what is achievable with classical resources. Most of the realistic metrological problems involve more than one unknown parameter, which led to the birth of multiparameter quantum metrology. In this context, a fundamental problem is to optimally allocate the finite available resources, depending on which parameters are treated as nuisance noises and which are the parameters of interest, in order to unlock wider regions of sub-SQL scaling. In this work, we accomplished both these tasks considering the scenario where the parameter of interest is either only the rotation angle or the angle and the fringes visibility. We experimentally showed that this approach is able to reach sub-SQL performances on the estimation precision even when unknown nuisance noises are present, for a resources range $O(30,000)$. The obtained results have shown the possibility of extending the advantages of multiparameter quantum metrology in the large resource domain. On the other, the methodology here described can find application in  large varieties of experimental platforms for quantum sensing with analogous noise models, thus representing a tool for future generations of quantum sensors, that can be employed to boost sensing scenarios like the dynamical tracking of biological reactions \cite{cimini2019adaptive}, but also for estimating the relative rotation of communicating stations \cite{krenn2015twisted}.
Finally, another key advantage of this approach is that it enables the system to be adaptable and resource-efficient, tailoring the estimation process to the specific characteristics and parameters of interest in various experimental setups, including more complex scenarios involving multiple entangled sensors \cite{brady2023entanglement}. This adaptability allows for improved accuracy and scalability, making it a valuable tool in the pursuit of ultra-precise sensing across diverse applications.

\begin{acknowledgments}

The Bayesian data analysis has been programmed with the Python framework PyTorch and ran on a GPU. The code can be found on github~\cite{github}. We gratefully acknowledge computational resources of the Center for High Performance Computing (CHPC) at SNS. This work is supported by the ERC Advanced grant PHOSPhOR (Photonics of Spin-Orbit Optical Phenomena; Grant Agreement No. 828978), by the Amaldi Research Center funded by the Ministero dell'Istruzione dell'Universit\`a e della Ricerca (Ministry of Education, University and Research) program ``Dipartimento di Eccellenza'' (CUP:B81I18001170001) and by MIUR (Ministero dell’Istruzione, dell’Università e della Ricerca) via project PRIN 2017 “Taming complexity via QUantum Strategies a Hybrid Integrated Photonic approach” (QUSHIP) Id. 2017SRNBRK.

\end{acknowledgments}

\clearpage

\appendix

\section{Derivation of multiparameter precision bounds}

\begin{figure}[!t]
	\includegraphics[width=\columnwidth]{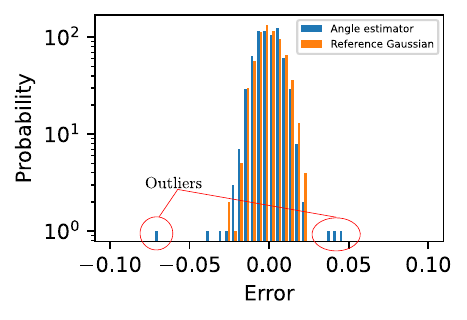}
	\caption{Probability density of the error for the rotation angle estimator (blue bars), compared with a reference Gaussian distributions (orange bars). This plot is for a estimation with $N=3000$ resource and a batchsize $b=3000$ (total number of runs). The outliers of the estimator distribution (which appear even further away from the central peak than the ones showed) are very impactful on the mean square error, while not affecting at all the median square error.}
	\label{fig:gaussian}
\end{figure}

In this section, we derive the multiparameter precision bound of Eq.~(2) in the main text. In order to do so, we  start from the quantity that we simulate, that are the measurements outcomes. The greedy algorithm selects for each photon consumed in the experiment the best value of $s$ among the available ones and the polarization basis $b$ that give us the maximum information gain. We called $\boldsymbol{r}_N$ the list of tuples containing these choices, together with the relative measurement outcome $o$ i.e.
\begin{equation}
    \boldsymbol{r}_N = \lbrace (s_1, b_1, o_1), (s_2, b_2, o_2), \cdots, (s_K, b_K, o_K) \rbrace \; ,
\end{equation}
Notice that the number of total resources used $N$ and the number of measurements $K$, i.e. the number of photons are different. We also call this string the trajectory of the estimation run. The most widely employed figure of merit for the precision of an estimator is the squared deviation from the true values of the parameters.
Given the estimators $\hat{\theta}^{(\boldsymbol{r}_N)}$ and $\hat{V}^{(\boldsymbol{r}_N)}_{s_i}$ for $\theta$ and $V_{s_i}$ respectively, and a weight matrix $G$ that codifies which parameters are of interests, we have introduced in the main text the error quantity
\begin{equation}
    \Delta^2_{\boldsymbol{r}_N, G} (\theta) := G_{1,1}|\hat{\theta}^{({\mathbf r}_N)}-\theta|^2 + \sum_{i=1}^4 G_{i+1\, , i+1} |\hat{V}^{({\mathbf r}_N)}_{s_i}-V_{s_i}|^2 \; ,
\end{equation}
its expectation value on the experimental run is $\Delta^2_G (\theta) := \mathbb{E}_{\boldsymbol{r}_N} [\Delta^2_{\boldsymbol{r}_N, G} (\theta)]$. We then take the expectation value of this precision on the prior distribution for $\theta$ and define $\Delta^2_G := \mathbb{E}_{\theta} [\Delta^2_{G} (\theta)]$. We can approximate these expressions by means of $M$ simulations for each of the $J$ angle $\theta_j$. So that we can write
\begin{equation}
    \Delta^2_G \simeq \frac{1}{MJ} \sum_{m=1}^{M} \sum_{j=1}^J \Delta^2_{\boldsymbol{r}_N^{m, j}, G} (\theta_j) \; , 
\end{equation}
where ${\boldsymbol{r}_N^{m, j}}$ is the trajectory of the $m$-th experimental run for the $j$-th angle. The figure of merit in Eq.~(1) of the main text is computed by taking the median of the $M$ quantities $\sum_{j=1}^J \Delta^2_{\boldsymbol{r}_N^{m, j}, G} (\theta_j)$ instead of the mean. We now see how the Cramèr-Rao (CR) bound sets a limit to $\Delta^2_G$ and how this can become a bound for the median error $\mathcal{M}_G^2$. Depending on the measurement basis chosen, the results $o_1, o_2 \in \lbrace -1, +1 \rbrace$ of the two polarization measurements for the $i-$th q-plate are distributed respectively according to 
\begin{equation}
\begin{split}
    & p_1(o_1 | s_1, \theta, V_{s_i}) := \frac{1}{2} \left(1 + o_1 \cdot V_{s_i} \cos 2 s_i \theta \right) \; \\ & p_2(o_2 | s_1, \theta, V_{s_i}) := \frac{1}{2} \left( 1 + o_2 \cdot V_{s_i} \sin 2 s_i \theta \right) \; .
    \label{eq:probabilities}
    \end{split}
\end{equation}
We have $\nu_i$ measurements for the q-plate $s_i$ in total, that we assume being evenly split between the two polarization basis. From these probabilities we can write the $5 \times 5$ Fisher information matrix $I$ (FI matrix) for the five parameters $(\theta, V_{s_1}, V_{s_2}, V_{s_3}, V_{s_4})$, whose non-zero elements are 
\begin{align*}
    I_{11} &= \sum_{i=1}^4 \frac{4 s_i^2 V_{s_i}^2 \nu_i \left( -4 + 3 V_{s_i}^2 + V_{s_i}^2 \cos 4 s_i \theta \right)}{-8 + 8 V_{s_i}^2 -V_{s_i}^4 +V_{s_i}^4 \cos 4 s_i \theta} \; ,\\
    I_{i+1, 1} &= -\frac{4 s_i V_{s_i}^3 \nu_i \cot 2 s_i \theta}{(V_{s_i}^2 - \csc ^2 s_i \theta)(V_{s_i}^2 -\sec ^2 s_i \theta)} \; , \\
    I_{i+1, i+1} &= 2 \nu_i \left( \frac{1}{-V_{s_i}^2 + \csc^2 s_i \theta} + \frac{1}{-V_{s_i}^2 + \sec^2 s_i \theta}\right) \; ,
\end{align*}
for $i=1, 2, 3, 4$, and $I_{i+1, 1} = I_{1, i+1}$ for symmetry. The Cramér-Rao bound, holding true for asymptotically unbiased estimators, is then expressed by the following inequality involving $\Delta^2_G (\theta)$:
\begin{equation}
    \Delta^2_G (\theta) \ge \text{Tr} \left( G \cdot I^{-1} \right) \; ,
\end{equation}
by taking the expectation value on the prior on $\theta$ we have
\begin{equation}
    \Delta^2_G = \mathbb{E}_\theta [\Delta^2_G (\theta)] \ge \mathbb{E}_\theta [\text{Tr} \left( G \cdot I^{-1} \right)] \ge \text{Tr} \left( G \cdot \mathbb{E}_\theta[I]^{-1} \right) \; .
    \label{eq:cr_bound_minimize}
\end{equation}
We now want to renormalize the uses of each q-plate $\nu_i$ in such a way to highlight the dependence on the total number of resources $N$, i.e. $\nu_i := x_i N$. The FI matrix $I$ becomes $I = N \widetilde{I}$, where the entries of $\widetilde{I}$ are similar to that of $I$, only that $\nu_i$ is substituted with $x_i$. The CR bound reads now
\begin{equation}
    \Delta^2_G \ge \frac{\text{Tr} \left( G \cdot \mathbb{E}_\theta[\widetilde{I}]^{-1} \right)}{N} \ge  \frac{C_G}{N}\; ,
\end{equation}
where the expectation value of the matrix $\widetilde{I}$ is diagonal with entries
\begin{eqnarray}
    \mathbb{E}_\theta [\widetilde{I}_{11}] &=& 4 \sum_{i=1}^4 x_i s_i^2 \left( 1 - \sqrt{1-V_{s_i}^2}\right) \; , \\
	\label{eq:averagePhase}
	\mathbb{E}_\theta [\widetilde{I}_{i+1, i+1}] &=& \frac{4 x_i \left( 1 - \sqrt{1-V_{s_i}^2} \right)}{V_{s_i}^2 \sqrt{1-V_{s_i}^2}}\, , \quad \text{for} \; i=1, \dots, 4 \; ,
	\label{eq:averageVisibilities}
\end{eqnarray}
and $C_G$ is the solution of the following minimization problem:
\begin{equation}
	\begin{cases}
	C_G = \min_{x_i} \text{Tr} \left( G \cdot \mathbb{E}_\theta [\widetilde{I}]^{-1} \right)\\
	\text{subject to} \; \sum_{i=1}^4 s_i x_i = 1 \\
	x_i \ge 0 
	\end{cases}\\
	\; .
\label{eq:semidefinite}
\end{equation}
In order to get a reference value for the median square error, we suppose that the estimators $\hat{\theta}^{(\boldsymbol{r}_N)}$ and $\hat{V}^{(\boldsymbol{r}_N)}_{s_i}$ are asymptotically normal and unbiased, we give some empirical evidence of this later in this section. Because the square of the deviations $|\hat{\theta}^{({\mathbf r}_N)}-\theta_j|^2$ and $|\hat{V}_{s_i}^{({\mathbf r}_N)}-V_{s_i}|^2$ are left-skewed and independent variables, we observe that
\begin{eqnarray}
\begin{split}
     \text{Median} \left[ \sum_{j=1}^J \Delta^2_{\boldsymbol{r}_N^{j, m}, G} (\theta_j) \right] & \ge  \sum_{j=1}^J \text{Median} [|\hat{\theta}^{({\mathbf r}_N^{m, j})}-\theta_j|^2] \\ &+ \text{Median} [|\hat{V}_{s_i}^{({\mathbf r}_N^{m, j})}-V_{s_i}|^2] \; .
\end{split}
\end{eqnarray}
The variable $\hat{\theta}^{({\mathbf r}_N)}-\theta_j$ is normally distributed and centred in zero. Under this hypothesis it is easy to show that the median of its square is proportional to the variance
\begin{equation}
	\text{Median} \left[ |\hat{\theta}^{({\mathbf r}_N)}-\theta|^2 \right]= \xi \mathbb{E}_{\boldsymbol{r}_N} [ |\hat{\theta}^{({\mathbf r}_N)}-\theta|^2 ] \; ,
	\label{eq:relationMedianMean}
\end{equation}
with a factor $\xi \simeq 0.4549$ that can be estimated numerically. Therefore, the bound on the median error of the estimation is
\begin{equation}
    \mathcal{M}^2_G \ge \frac{\xi C_G}{N} := \frac{\widetilde{C_G}}{N}\; ,
	\label{eq:medianCramer}
\end{equation}
with $\widetilde{C_G}=\xi C_G$. We will now briefly prove the relation between the median and the mean square error. Under the aforementioned assumptions we have $\hat{\theta}^{({\mathbf r}_N)}-\theta \sim \mathcal{N} \left(0, \sigma^2 \right)$. From the formula for the transformation of the probability distributions we compute the probability density function for $y:=|\hat{\theta}^{({\mathbf r}_N)}-\theta|^2$, which is
\begin{equation}
f(y) = \frac{1}{\sigma \sqrt{2 \pi y}} e^{-\frac{y}{\sigma^2}} \; .
\end{equation}
The mean value of $y$ is by definition $\sigma^2$, i.e.,
\begin{equation}
\int_{0}^\infty \frac{y}{\sigma \sqrt{2 \pi y}} e^{-\frac{1}{2} \frac{y}{\sigma^2}} d y = \sigma^2 \, .
\end{equation}
The median $M := \text{Median} | \hat{\theta}_j - \theta_j |^2 = \text{Median}(y)$ is defined by the implicit formula
\begin{equation}
\int_{0}^M \frac{1}{\sigma \sqrt{2 \pi y}} e^{-\frac{1}{2} \frac{y}{\sigma^2}} d y = \frac{1}{2} \; .
\end{equation}
By introducing $k=\frac{y}{\sigma^2}$ and $M = \xi \sigma^2$ and changing variable in the above integral we get
\begin{equation}
\int_{0}^{\xi} \frac{1}{\sqrt{2 \pi k}} e^{-\frac{k}{2}} d k = \frac{1}{2} \; .
\end{equation}
In this integral the only unknown is $\xi$, and it can be solved numerically to get Eq.~\eqref{eq:relationMedianMean}, since by definition $xi$ is the proportionality factor between the median and the variance.
In Fig.~\ref{fig:gaussian} we report the comparison of the estimator for the rotation angle produced by the Bayesian procedure with a reference Gaussian. This justifies empirically the affirmation we made in the paper that the distribution for the estimator is asymptotically a bias-less Gaussian affected by outliers. What we measure with the median square error is the width of the central Gaussian-like part of the blue distribution, while ignoring the outliers.

\section{The Bayesian algorithm}
In this section we complement the presentation of the Bayesian algorithm that we gave in the main text, with the application to our q-plates setup in mind. With respect to the original formulation, we made a few corrections necessary because of the circular nature of the angular variable that we are going to measure. In every experiment, $n_p = 5000$ particles have been used. The parameters to estimate are collected in the vector $\boldsymbol{x} := \left( \theta, V_{s_1}, V_{s_2}, V_{s_3}, V_{s_4} \right)$, that contains the phase in the first entry and the four visibilities in the other ones. The Granade's approach~\cite{granade2012robust} is based on a particle filter, which represents on a computer the posterior probability distribution with the ensemble $\mathcal{E} := \lbrace (\boldsymbol{x}^k, w^k) \rbrace$, of $n_p$ pairs composed of a particle $k$ in the position $\boldsymbol{x}^k$ and with a weight $w_i$.

The weights are updated after each measurements with the Bayes rule, i.e.
\begin{equation}
	w^{k}_{t+1} \propto w_t^k p(o_t|s_i, \boldsymbol{x}^k) \; ,
	\label{eq:BayesUpdate}
\end{equation}
and depend therefore on the measurement step $t$. The probability $p(o_t|s_i, \boldsymbol{x}^k)$ is either $p_1$ or $p_2$ according to the basis of the polarization measurement chosen. The weights $w_{t+1}^k$ are renormalized after each update. In the following, we will drop the dependence on the measurement step $t$ except when needed. The $j$-th component of the $k$-th particle of the ensemble will be represented as $x_j^k$, and could correspond to the phase if $j=0$, that is $x^k_0 = \theta^k$ or to one of the visibilities if $j=1, 2, 3, 4$, that is $x^k_j = V^k_{s_j}$. The mean of the angular values is computed as
\begin{equation}
	\hat{\mu}_0 := \arg \left[ \sum_{k=1}^{n_{p}} w^k \exp \left( i \theta^k \right) \right] \; ,
\end{equation}
while the mean values of the visibilities are
\begin{equation}
	\hat{\mu}_j = \sum_{k=1}^{n_p} w^k V^k_{s_j} \; .
\end{equation}
Together they form the vectorial mean of the posterior distribution $\boldsymbol{\hat{\mu}} = (\hat{\mu}_0, \hat{\mu}_1, \hat{\mu}_2, \hat{\mu}_3, \hat{\mu}_4)$. Its components will also be our estimators for the parameters i.e. $\hat{\theta}:=\hat{\mu}_0$ and $\hat{V}_{s_i}:=\hat{\mu}_i$. The covariance matrix is defined as
\begin{equation}
	\text{Cov}_{ij} := \sum_{k=1}^{n_{p}} w^k d_i(x^k_i, \hat{\mu}_i)  d_j(x^k_j, \hat{\mu}_j) \; ,
\end{equation}

where for $i, j=1$ we use the circular distance, i.e.
\begin{equation}
	d_1(x^k_1, \hat{\mu}_1) = \pi - | (x^k_1 - \hat{\mu}_1) \mod 2 \pi - \pi| \; ,
\end{equation}
while for $i, j > 4$ we use the regular distance $d_i = |x^k_i - \hat{\mu}_i|$. The scalar variance of the posterior distribution is
\begin{equation}
	\sigma^2 := \text{Tr} \left[ G \cdot \text{Cov} \right] = \sum_{i, j} G_{ij} \text{Cov}_{ij} \; ,
\end{equation}
by simulating each one of the possible $8$ experiments ($4$ q-plates and $2$ polarization basis among which we have to choose) and selecting the one with the lowest expected variance we realize the greedy optimization described in the main text. In particular, we compute for each combination of q-plate and polarization basis the expected scalar variance, i.e.
\begin{equation}
	\mathbb{E} \left[ \sigma^2 \right] := \sum_{o=\pm1} p_{1/2}(o|s_i, \hat{V}_{s_i}, \hat{\theta}_j) \sigma^2 (o)\; ,
\end{equation}
where $\sigma^2(o)$ is the variance of the posterior distribution where the weigh update of Eq.~\eqref{eq:BayesUpdate} is performed assuming the outcome $o$ for the measurement. This variance is weighted with an estimation for the probability of getting the outcome $o$. In this way the algorithm attempts to concentrate as much as possible the distribution around its mean, without however planning for more than one step ahead in the future.

The simplest possible non-greedy extension of this would be to simulate two measurement steps, this would mean computing $64$ possible expected variances. Simulating many more steps becomes quickly unfeasible,  and a more refined approach, possibly involving machine learning is required. The resampling strategy of the Granade procedure in~\cite{granade2012robust} has also undergone minor changes to adapt it to the phase estimation problem.

\section{Non-adaptive offline strategy for the single parameter estimation} 

For clarity we report here a very brief comparison with the single parameter estimation strategy of~\cite{cimini2023experimental}. In the previous experiment, the apparatus is the same, but there we are doing single parameter estimation. The visibilities are measured beforehand and are used (together with the charges of the available q-plates) to compute an optimal strategy, that is the optimal use of q-plates. In that situation at contrary to this article, the strategy is non-adaptive and computed off-line. The online strategy of this paper is useful when the visibilities are not known in advance and can change during the experiment.

\section{Values of the angles and the visibilities}
\label{sec:values}
The value of the rotation angle is known from the mechanical platform on which the receiving end of the apparatus is mounted and the visibilities are computed from the estimator
\begin{equation}
    \hat{V} = \sqrt{\frac{\nu \left[ \left(2 f_0 -1 \right)^2 + (2 f_+ -1)^2 \right] -1}{\nu -1}} \; ,
\end{equation}
where $f_0$ and $f_+$ are the frequencies of the outcomes $o=0$ and $o=+$ of the two polarization measurements for a fixed phase and q-plate, and $\nu$ is the number of experiments executed for each polarization. These estimators for the visibilities are evaluated prior and independently of the Bayesian procedure. The results will be considered the ``true'' values of the visibilities in evaluating the precision of the Bayesian approach, i.e. $\Delta_{{\boldsymbol{r}_n}, G}^2 (\theta_j)$. These are

\begin{center}
    \begin{tabular}{ |c|c|c|c|c| } 
    \hline
    $\theta$ & $V_1$ & $V_2$ & $V_3$ & $V_4$\\
    \hline
    0.00235 & 0.8776 & 0.9091 & 0.8445 & 0.7038\\ 0.06145 & 0.9085 & 0.8934 & 0.8007 & 0.7611\\
    0.38000 & 0.9399 & 0.9153 & 0.7936 & 0.7222\\
    0.49620 & 0.9211 & 0.9315 & 0.7261 & 0.8186\\
    1.6645 & 0.9331 & 0.8914 & 0.8691 & 0.7312\\
    1.8750 & 0.9599 & 0.9081 & 0.8762 & 0.6618\\
    2.5900 & 0.9187 & 0.9587 & 0.8775 & 0.6848\\
    2.9600 & 0.8986 & 0.9321 & 0.8700 & 0.7528\\
    - & 0.9197 & 0.9174 & 0.8322 & 0.7295\\
    \hline
    \end{tabular}
    \label{tab:visibilities}
\end{center}
The above table contains the values of the eight angles analyzed in the experiment and their corresponding visibilities for each q-plate configuration. The last line reports the mean values of the visibilities. We did not stress this point in the main text, but for each angle we have a slightly different value of the visibility, and this is taken into account in computing the square error.


\begin{thebibliography}{0}%
\makeatletter
\providecommand \@ifxundefined [1]{%
 \@ifx{#1\undefined}
}%
\providecommand \@ifnum [1]{%
 \ifnum #1\expandafter \@firstoftwo
 \else \expandafter \@secondoftwo
 \fi
}%
\providecommand \@ifx [1]{%
 \ifx #1\expandafter \@firstoftwo
 \else \expandafter \@secondoftwo
 \fi
}%
\providecommand \natexlab [1]{#1}%
\providecommand \enquote  [1]{``#1''}%
\providecommand \bibnamefont  [1]{#1}%
\providecommand \bibfnamefont [1]{#1}%
\providecommand \citenamefont [1]{#1}%
\providecommand \href@noop [0]{\@secondoftwo}%
\providecommand \href [0]{\begingroup \@sanitize@url \@href}%
\providecommand \@href[1]{\@@startlink{#1}\@@href}%
\providecommand \@@href[1]{\endgroup#1\@@endlink}%
\providecommand \@sanitize@url [0]{\catcode `\\12\catcode `\$12\catcode
  `\&12\catcode `\#12\catcode `\^12\catcode `\_12\catcode `\%12\relax}%
\providecommand \@@startlink[1]{}%
\providecommand \@@endlink[0]{}%
\providecommand \url  [0]{\begingroup\@sanitize@url \@url }%
\providecommand \@url [1]{\endgroup\@href {#1}{\urlprefix }}%
\providecommand \urlprefix  [0]{URL }%
\providecommand \Eprint [0]{\href }%
\providecommand \doibase [0]{http://dx.doi.org/}%
\providecommand \selectlanguage [0]{\@gobble}%
\providecommand \bibinfo  [0]{\@secondoftwo}%
\providecommand \bibfield  [0]{\@secondoftwo}%
\providecommand \translation [1]{[#1]}%
\providecommand \BibitemOpen [0]{}%
\providecommand \bibitemStop [0]{}%
\providecommand \bibitemNoStop [0]{.\EOS\space}%
\providecommand \EOS [0]{\spacefactor3000\relax}%
\providecommand \BibitemShut  [1]{\csname bibitem#1\endcsname}%
\let\auto@bib@innerbib\@empty
\end{thebibliography}%


\begin{thebibliography}{10}
\expandafter\ifx\csname url\endcsname\relax
  \def\url#1{\texttt{#1}}\fi
\expandafter\ifx\csname urlprefix\endcsname\relax\def\urlprefix{URL }\fi
\providecommand{\bibinfo}[2]{#2}
\providecommand{\eprint}[2][]{\url{#2}}

\bibitem{Giovannetti1330}
\bibinfo{author}{Giovannetti, V.}, \bibinfo{author}{Lloyd, S.} \&
  \bibinfo{author}{Maccone, L.}
\newblock \bibinfo{title}{Quantum-enhanced measurements: Beating the standard
  quantum limit}.
\newblock \emph{\bibinfo{journal}{Science}} \textbf{\bibinfo{volume}{306}},
  \bibinfo{pages}{1330--1336} (\bibinfo{year}{2004}).
\newblock \urlprefix\url{https://science.sciencemag.org/content/306/5700/1330}.

\bibitem{PhysRevLett.96.010401}
\bibinfo{author}{Giovannetti, V.}, \bibinfo{author}{Lloyd, S.} \&
  \bibinfo{author}{Maccone, L.}
\newblock \bibinfo{title}{Quantum metrology}.
\newblock \emph{\bibinfo{journal}{Phys. Rev. Lett.}}
  \textbf{\bibinfo{volume}{96}}, \bibinfo{pages}{010401}
  (\bibinfo{year}{2006}).
\newblock
  \urlprefix\url{https://link.aps.org/doi/10.1103/PhysRevLett.96.010401}.

\bibitem{barbieri2022optical}
\bibinfo{author}{Barbieri, M.}
\newblock \bibinfo{title}{Optical quantum metrology}.
\newblock \emph{\bibinfo{journal}{PRX Quantum}} \textbf{\bibinfo{volume}{3}},
  \bibinfo{pages}{010202} (\bibinfo{year}{2022}).

\bibitem{Giovannetti}
\bibinfo{author}{Giovannetti, V.}, \bibinfo{author}{Lloyd, S.} \&
  \bibinfo{author}{Maccone, L.}
\newblock \bibinfo{title}{Advances in quantum metrology}.
\newblock \emph{\bibinfo{journal}{Nat. Photonics}}
  \textbf{\bibinfo{volume}{5}}, \bibinfo{pages}{222--229}
  (\bibinfo{year}{2011}).
\newblock \urlprefix\url{https://doi.org/10.1038/nphoton.2011.35}.

\bibitem{avsreview2020}
\bibinfo{author}{Polino, E.}, \bibinfo{author}{Valeri, M.},
  \bibinfo{author}{Spagnolo, N.} \& \bibinfo{author}{Sciarrino, F.}
\newblock \bibinfo{title}{Photonic quantum metrology}.
\newblock \emph{\bibinfo{journal}{AVS Quantum Science}}
  \textbf{\bibinfo{volume}{2}}, \bibinfo{pages}{024703} (\bibinfo{year}{2020}).

\bibitem{liu2019quantum}
\bibinfo{author}{Liu, J.}, \bibinfo{author}{Yuan, H.}, \bibinfo{author}{Lu,
  X.-M.} \& \bibinfo{author}{Wang, X.}
\newblock \bibinfo{title}{Quantum fisher information matrix and multiparameter
  estimation}.
\newblock \emph{\bibinfo{journal}{Journal of Physics A: Mathematical and
  Theoretical}} \textbf{\bibinfo{volume}{53}}, \bibinfo{pages}{023001}
  (\bibinfo{year}{2019}).

\bibitem{suzuki2020quantum}
\bibinfo{author}{Suzuki, J.}, \bibinfo{author}{Yang, Y.} \&
  \bibinfo{author}{Hayashi, M.}
\newblock \bibinfo{title}{Quantum state estimation with nuisance parameters}.
\newblock \emph{\bibinfo{journal}{Journal of Physics A: Mathematical and
  Theoretical}} \textbf{\bibinfo{volume}{53}}, \bibinfo{pages}{453001}
  (\bibinfo{year}{2020}).

\bibitem{suzuki2020nuisance}
\bibinfo{author}{Suzuki, J.}
\newblock \bibinfo{title}{Nuisance parameter problem in quantum estimation
  theory: Tradeoff relation and qubit examples}.
\newblock \emph{\bibinfo{journal}{Journal of Physics A: Mathematical and
  Theoretical}} \textbf{\bibinfo{volume}{53}}, \bibinfo{pages}{264001}
  (\bibinfo{year}{2020}).

\bibitem{multiparameter_QM}
\bibinfo{author}{Magdalena}, \bibinfo{author}{Baumgratz, T.} \&
  \bibinfo{author}{Datta, A.}
\newblock \bibinfo{title}{Multi-parameter quantum metrology}.
\newblock \emph{\bibinfo{journal}{Advances in Physics: X}}
  \textbf{\bibinfo{volume}{1}}, \bibinfo{pages}{621--639}
  (\bibinfo{year}{2016}).
\newblock \urlprefix\url{https://doi.org/10.1080/23746149.2016.1230476}.
\newblock \eprint{https://doi.org/10.1080/23746149.2016.1230476}.

\bibitem{albarelli2020perspective}
\bibinfo{author}{Albarelli, F.}, \bibinfo{author}{Barbieri, M.},
  \bibinfo{author}{Genoni, M.~G.} \& \bibinfo{author}{Gianani, I.}
\newblock \bibinfo{title}{A perspective on multiparameter quantum metrology:
  from theoretical tools to applications in quantum imaging}.
\newblock \emph{\bibinfo{journal}{Physics Letters A}}
  \textbf{\bibinfo{volume}{384}}, \bibinfo{pages}{126311}
  (\bibinfo{year}{2020}).

\bibitem{polino2019experimental}
\bibinfo{author}{Polino, E.} \emph{et~al.}
\newblock \bibinfo{title}{Experimental multiphase estimation on a chip}.
\newblock \emph{\bibinfo{journal}{Optica}} \textbf{\bibinfo{volume}{6}},
  \bibinfo{pages}{288--295} (\bibinfo{year}{2019}).

\bibitem{liu2021distributed}
\bibinfo{author}{Liu, L.-Z.} \emph{et~al.}
\newblock \bibinfo{title}{Distributed quantum phase estimation with entangled
  photons}.
\newblock \emph{\bibinfo{journal}{Nature Photonics}}
  \textbf{\bibinfo{volume}{15}}, \bibinfo{pages}{137--142}
  (\bibinfo{year}{2021}).

\bibitem{hong2021quantum}
\bibinfo{author}{Hong, S.} \emph{et~al.}
\newblock \bibinfo{title}{Quantum enhanced multiple-phase estimation with
  multi-mode n00n states}.
\newblock \emph{\bibinfo{journal}{Nature communications}}
  \textbf{\bibinfo{volume}{12}}, \bibinfo{pages}{1--8} (\bibinfo{year}{2021}).

\bibitem{valeri2023experimental}
\bibinfo{author}{Valeri, M.} \emph{et~al.}
\newblock \bibinfo{title}{Experimental multiparameter quantum metrology in
  adaptive regime}.
\newblock \emph{\bibinfo{journal}{Physical Review Research}}
  \textbf{\bibinfo{volume}{5}}, \bibinfo{pages}{013138} (\bibinfo{year}{2023}).

\bibitem{cimini2022deep}
\bibinfo{author}{Cimini, V.} \emph{et~al.}
\newblock \bibinfo{title}{Deep reinforcement learning for quantum
  multiparameter estimation}.
\newblock \emph{\bibinfo{journal}{arXiv preprint arXiv:2209.00671}}
  (\bibinfo{year}{2022}).

\bibitem{humphreys2013quantum}
\bibinfo{author}{Humphreys, P.~C.}, \bibinfo{author}{Barbieri, M.},
  \bibinfo{author}{Datta, A.} \& \bibinfo{author}{Walmsley, I.~A.}
\newblock \bibinfo{title}{Quantum enhanced multiple phase estimation}.
\newblock \emph{\bibinfo{journal}{Physical Review Letters}}
  \textbf{\bibinfo{volume}{111}}, \bibinfo{pages}{070403}
  (\bibinfo{year}{2013}).

\bibitem{PhysRevLett.128.040504}
\bibinfo{author}{G\'orecki, W.} \&
  \bibinfo{author}{Demkowicz-Dobrza\ifmmode~\acute{n}\else \'{n}\fi{}ski, R.}
\newblock \bibinfo{title}{Multiple-phase quantum interferometry: Real and
  apparent gains of measuring all the phases simultaneously}.
\newblock \emph{\bibinfo{journal}{Phys. Rev. Lett.}}
  \textbf{\bibinfo{volume}{128}}, \bibinfo{pages}{040504}
  (\bibinfo{year}{2022}).
\newblock
  \urlprefix\url{https://link.aps.org/doi/10.1103/PhysRevLett.128.040504}.

\bibitem{belliardo2021incompatibility}
\bibinfo{author}{Belliardo, F.} \& \bibinfo{author}{Giovannetti, V.}
\newblock \bibinfo{title}{Incompatibility in quantum parameter estimation}.
\newblock \emph{\bibinfo{journal}{New Journal of Physics}}
  \textbf{\bibinfo{volume}{23}}, \bibinfo{pages}{063055}
  (\bibinfo{year}{2021}).

\bibitem{Yue}
\bibinfo{author}{Yue, J.-D.}, \bibinfo{author}{Zhang, Y.-R.} \&
  \bibinfo{author}{Fan, H.}
\newblock \bibinfo{title}{Quantum-enhanced metrology for multiple phase
  estimation with noise}.
\newblock \emph{\bibinfo{journal}{Scientific Reports}}
  \textbf{\bibinfo{volume}{4}} (\bibinfo{year}{2014}).

\bibitem{Roccia:18}
\bibinfo{author}{Roccia, E.} \emph{et~al.}
\newblock \bibinfo{title}{Multiparameter approach to quantum phase estimation
  with limited visibility}.
\newblock \emph{\bibinfo{journal}{Optica}} \textbf{\bibinfo{volume}{5}},
  \bibinfo{pages}{1171--1176} (\bibinfo{year}{2018}).
\newblock
  \urlprefix\url{http://www.osapublishing.org/optica/abstract.cfm?URI=optica-5-10-1171}.

\bibitem{roccia2017entangling}
\bibinfo{author}{Roccia, E.} \emph{et~al.}
\newblock \bibinfo{title}{Entangling measurements for multiparameter estimation
  with two qubits}.
\newblock \emph{\bibinfo{journal}{Quantum Science and Technology}}
  \textbf{\bibinfo{volume}{3}}, \bibinfo{pages}{01LT01} (\bibinfo{year}{2017}).

\bibitem{cimini2019quantum}
\bibinfo{author}{Cimini, V.} \emph{et~al.}
\newblock \bibinfo{title}{Quantum sensing for dynamical tracking of chemical
  processes}.
\newblock \emph{\bibinfo{journal}{Physical Review A}}
  \textbf{\bibinfo{volume}{99}}, \bibinfo{pages}{053817}
  (\bibinfo{year}{2019}).

\bibitem{cimini2019adaptive}
\bibinfo{author}{Cimini, V.} \emph{et~al.}
\newblock \bibinfo{title}{Adaptive tracking of enzymatic reactions with quantum
  light}.
\newblock \emph{\bibinfo{journal}{Optics Express}}
  \textbf{\bibinfo{volume}{27}}, \bibinfo{pages}{35245--35256}
  (\bibinfo{year}{2019}).

\bibitem{Vidrighin}
\bibinfo{author}{Vidrighin, M.~D.} \emph{et~al.}
\newblock \bibinfo{title}{Joint estimation of phase and phase diffusion for
  quantum metrology}.
\newblock \emph{\bibinfo{journal}{Nature Communications}}
  \textbf{\bibinfo{volume}{5}} (\bibinfo{year}{2014}).

\bibitem{PhysRevLett.106.153603}
\bibinfo{author}{Genoni, M.~G.}, \bibinfo{author}{Olivares, S.} \&
  \bibinfo{author}{Paris, M. G.~A.}
\newblock \bibinfo{title}{Optical phase estimation in the presence of phase
  diffusion}.
\newblock \emph{\bibinfo{journal}{Phys. Rev. Lett.}}
  \textbf{\bibinfo{volume}{106}}, \bibinfo{pages}{153603}
  (\bibinfo{year}{2011}).
\newblock
  \urlprefix\url{https://link.aps.org/doi/10.1103/PhysRevLett.106.153603}.

\bibitem{PhysRevA.84.012103}
\bibinfo{author}{Matsuzaki, Y.}, \bibinfo{author}{Benjamin, S.~C.} \&
  \bibinfo{author}{Fitzsimons, J.}
\newblock \bibinfo{title}{Magnetic field sensing beyond the standard quantum
  limit under the effect of decoherence}.
\newblock \emph{\bibinfo{journal}{Phys. Rev. A}} \textbf{\bibinfo{volume}{84}},
  \bibinfo{pages}{012103} (\bibinfo{year}{2011}).
\newblock \urlprefix\url{https://link.aps.org/doi/10.1103/PhysRevA.84.012103}.

\bibitem{goldberg2020multiphase}
\bibinfo{author}{Goldberg, A.~Z.} \emph{et~al.}
\newblock \bibinfo{title}{Multiphase estimation without a reference mode}.
\newblock \emph{\bibinfo{journal}{Physical Review A}}
  \textbf{\bibinfo{volume}{102}}, \bibinfo{pages}{022230}
  (\bibinfo{year}{2020}).

\bibitem{rubio2020quantum}
\bibinfo{author}{Rubio, J.}, \bibinfo{author}{Knott, P.~A.},
  \bibinfo{author}{Proctor, T.~J.} \& \bibinfo{author}{Dunningham, J.~A.}
\newblock \bibinfo{title}{Quantum sensing networks for the estimation of linear
  functions}.
\newblock \emph{\bibinfo{journal}{Journal of Physics A: Mathematical and
  Theoretical}} \textbf{\bibinfo{volume}{53}}, \bibinfo{pages}{344001}
  (\bibinfo{year}{2020}).

\bibitem{rubio2019limited}
\bibinfo{author}{Rubio, J.} \& \bibinfo{author}{Dunningham, J.}
\newblock \bibinfo{title}{Quantum metrology in the presence of limited data}.
\newblock \emph{\bibinfo{journal}{New Journal of Physics}}
  \textbf{\bibinfo{volume}{21}}, \bibinfo{pages}{043037}
  (\bibinfo{year}{2019}).

\bibitem{rubio2020bayesian}
\bibinfo{author}{Rubio, J.} \& \bibinfo{author}{Dunningham, J.}
\newblock \bibinfo{title}{Bayesian multiparameter quantum metrology with
  limited data}.
\newblock \emph{\bibinfo{journal}{Physical Review A}}
  \textbf{\bibinfo{volume}{101}}, \bibinfo{pages}{032114}
  (\bibinfo{year}{2020}).

\bibitem{wiseman1995adaptive}
\bibinfo{author}{Wiseman, H.~M.}
\newblock \bibinfo{title}{Adaptive phase measurements of optical modes: Going
  beyond the marginal q distribution}.
\newblock \emph{\bibinfo{journal}{Physical Review Letters}}
  \textbf{\bibinfo{volume}{75}}, \bibinfo{pages}{4587} (\bibinfo{year}{1995}).

\bibitem{berry2000optimal}
\bibinfo{author}{Berry, D.~W.} \& \bibinfo{author}{Wiseman, H.~M.}
\newblock \bibinfo{title}{Optimal states and almost optimal adaptive
  measurements for quantum interferometry}.
\newblock \emph{\bibinfo{journal}{Physical review letters}}
  \textbf{\bibinfo{volume}{85}}, \bibinfo{pages}{5098} (\bibinfo{year}{2000}).

\bibitem{berni2015ab}
\bibinfo{author}{Berni, A.~A.} \emph{et~al.}
\newblock \bibinfo{title}{Ab initio quantum-enhanced optical phase estimation
  using real-time feedback control}.
\newblock \emph{\bibinfo{journal}{Nature Photonics}}
  \textbf{\bibinfo{volume}{9}}, \bibinfo{pages}{577} (\bibinfo{year}{2015}).

\bibitem{Cimini:19}
\bibinfo{author}{Cimini, V.} \emph{et~al.}
\newblock \bibinfo{title}{Adaptive tracking of enzymatic reactions with quantum
  light}.
\newblock \emph{\bibinfo{journal}{Opt. Express}} \textbf{\bibinfo{volume}{27}},
  \bibinfo{pages}{35245--35256} (\bibinfo{year}{2019}).
\newblock
  \urlprefix\url{http://opg.optica.org/oe/abstract.cfm?URI=oe-27-24-35245}.

\bibitem{hentschel2011efficient}
\bibinfo{author}{Hentschel, A.} \& \bibinfo{author}{Sanders, B.~C.}
\newblock \bibinfo{title}{Efficient algorithm for optimizing adaptive quantum
  metrology processes}.
\newblock \emph{\bibinfo{journal}{Physical Review Letters}}
  \textbf{\bibinfo{volume}{107}}, \bibinfo{pages}{233601}
  (\bibinfo{year}{2011}).

\bibitem{lovett2013differential}
\bibinfo{author}{Lovett, N.~B.}, \bibinfo{author}{Crosnier, C.},
  \bibinfo{author}{Perarnau-Llobet, M.} \& \bibinfo{author}{Sanders, B.~C.}
\newblock \bibinfo{title}{Differential evolution for many-particle adaptive
  quantum metrology}.
\newblock \emph{\bibinfo{journal}{Physical Review Letters}}
  \textbf{\bibinfo{volume}{110}}, \bibinfo{pages}{220501}
  (\bibinfo{year}{2013}).

\bibitem{hentschel2009adaptive}
\bibinfo{author}{Hentschel, A.} \& \bibinfo{author}{Sanders, B.~C.}
\newblock \bibinfo{title}{Machine learning for precise quantum measurement}.
\newblock \emph{\bibinfo{journal}{Physical Review Letters}}
  \textbf{\bibinfo{volume}{104}}, \bibinfo{pages}{063603}
  (\bibinfo{year}{2009}).

\bibitem{piccoloLume}
\bibinfo{author}{Lumino, A.} \emph{et~al.}
\newblock \bibinfo{title}{Experimental phase estimation enhanced by machine
  learning}.
\newblock \emph{\bibinfo{journal}{Physical Review Applied}}
  \textbf{\bibinfo{volume}{10}}, \bibinfo{pages}{044033}
  (\bibinfo{year}{2018}).

\bibitem{rambhatla2020adaptive}
\bibinfo{author}{Rambhatla, K.} \emph{et~al.}
\newblock \bibinfo{title}{Adaptive phase estimation through a genetic
  algorithm}.
\newblock \emph{\bibinfo{journal}{Physical Review Research}}
  \textbf{\bibinfo{volume}{2}}, \bibinfo{pages}{033078} (\bibinfo{year}{2020}).

\bibitem{cimini2023deep}
\bibinfo{author}{Cimini, V.} \emph{et~al.}
\newblock \bibinfo{title}{Deep reinforcement learning for quantum
  multiparameter estimation}.
\newblock \emph{\bibinfo{journal}{Advanced Photonics}}
  \textbf{\bibinfo{volume}{5}}, \bibinfo{pages}{016005} (\bibinfo{year}{2023}).

\bibitem{armen2002adaptive}
\bibinfo{author}{Armen, M.~A.}, \bibinfo{author}{Au, J.~K.},
  \bibinfo{author}{Stockton, J.~K.}, \bibinfo{author}{Doherty, A.~C.} \&
  \bibinfo{author}{Mabuchi, H.}
\newblock \bibinfo{title}{Adaptive homodyne measurement of optical phase}.
\newblock \emph{\bibinfo{journal}{Physical Review Letters}}
  \textbf{\bibinfo{volume}{89}}, \bibinfo{pages}{133602}
  (\bibinfo{year}{2002}).

\bibitem{higgins2007entanglement}
\bibinfo{author}{Higgins, B.~L.}, \bibinfo{author}{Berry, D.~W.},
  \bibinfo{author}{Bartlett, S.~D.}, \bibinfo{author}{Wiseman, H.~M.} \&
  \bibinfo{author}{Pryde, G.~J.}
\newblock \bibinfo{title}{Entanglement-free heisenberg-limited phase
  estimation}.
\newblock \emph{\bibinfo{journal}{Nature}} \textbf{\bibinfo{volume}{450}},
  \bibinfo{pages}{393--396} (\bibinfo{year}{2007}).

\bibitem{daryanoosh2018experimental}
\bibinfo{author}{Daryanoosh, S.}, \bibinfo{author}{Slussarenko, S.},
  \bibinfo{author}{Berry, D.~W.}, \bibinfo{author}{Wiseman, H.~M.} \&
  \bibinfo{author}{Pryde, G.~J.}
\newblock \bibinfo{title}{Experimental optical phase measurement approaching
  the exact heisenberg limit}.
\newblock \emph{\bibinfo{journal}{Nature Communications}}
  \textbf{\bibinfo{volume}{9}}, \bibinfo{pages}{4606} (\bibinfo{year}{2018}).

\bibitem{wheatley2010adaptive}
\bibinfo{author}{Wheatley, T.} \emph{et~al.}
\newblock \bibinfo{title}{Adaptive optical phase estimation using
  time-symmetric quantum smoothing}.
\newblock \emph{\bibinfo{journal}{Physical Review Letters}}
  \textbf{\bibinfo{volume}{104}}, \bibinfo{pages}{093601}
  (\bibinfo{year}{2010}).

\bibitem{Valeri2020}
\bibinfo{author}{Valeri, M.} \emph{et~al.}
\newblock \bibinfo{title}{Experimental adaptive bayesian estimation of multiple
  phases with limited data}.
\newblock \emph{\bibinfo{journal}{npj Quantum Information}}
  \textbf{\bibinfo{volume}{6}}, \bibinfo{pages}{92} (\bibinfo{year}{2020}).
\newblock \urlprefix\url{https://doi.org/10.1038/s41534-020-00326-6}.

\bibitem{tse2019quantum}
\bibinfo{author}{Tse, M.~e.} \emph{et~al.}
\newblock \bibinfo{title}{Quantum-enhanced advanced ligo detectors in the era
  of gravitational-wave astronomy}.
\newblock \emph{\bibinfo{journal}{Physical Review Letters}}
  \textbf{\bibinfo{volume}{123}}, \bibinfo{pages}{231107}
  (\bibinfo{year}{2019}).

\bibitem{guo2020distributed}
\bibinfo{author}{Guo, X.} \emph{et~al.}
\newblock \bibinfo{title}{Distributed quantum sensing in a continuous-variable
  entangled network}.
\newblock \emph{\bibinfo{journal}{Nature Physics}}
  \textbf{\bibinfo{volume}{16}}, \bibinfo{pages}{281--284}
  (\bibinfo{year}{2020}).

\bibitem{xia2020demonstration}
\bibinfo{author}{Xia, Y.} \emph{et~al.}
\newblock \bibinfo{title}{Demonstration of a reconfigurable entangled
  radio-frequency photonic sensor network}.
\newblock \emph{\bibinfo{journal}{Physical Review Letters}}
  \textbf{\bibinfo{volume}{124}}, \bibinfo{pages}{150502}
  (\bibinfo{year}{2020}).

\bibitem{goldberg2018quantum}
\bibinfo{author}{Goldberg, A.~Z.} \& \bibinfo{author}{James, D.~F.}
\newblock \bibinfo{title}{Quantum-limited euler angle measurements using
  anticoherent states}.
\newblock \emph{\bibinfo{journal}{Physical Review A}}
  \textbf{\bibinfo{volume}{98}}, \bibinfo{pages}{032113}
  (\bibinfo{year}{2018}).

\bibitem{goldberg2021rotation}
\bibinfo{author}{Goldberg, A.~Z.}, \bibinfo{author}{Klimov, A.~B.},
  \bibinfo{author}{Leuchs, G.} \& \bibinfo{author}{S{\'a}nchez-Soto, L.~L.}
\newblock \bibinfo{title}{Rotation sensing at the ultimate limit}.
\newblock \emph{\bibinfo{journal}{Journal of Physics: Photonics}}
  \textbf{\bibinfo{volume}{3}}, \bibinfo{pages}{022008} (\bibinfo{year}{2021}).

\bibitem{dambrosio_gear2013}
\bibinfo{author}{D'Ambrosio, V.} \emph{et~al.}
\newblock \bibinfo{title}{Photonic polarization gears for ultra-sensitive
  angular measurements}.
\newblock \emph{\bibinfo{journal}{Nat. Comm.}} \textbf{\bibinfo{volume}{4}},
  \bibinfo{pages}{2432} (\bibinfo{year}{2013}).
\newblock \urlprefix\url{https://www.nature.com/articles/ncomms3432}.

\bibitem{cimini2023experimental}
\bibinfo{author}{Cimini, V.} \emph{et~al.}
\newblock \bibinfo{title}{Experimental metrology beyond the standard quantum
  limit for a wide resources range}.
\newblock \emph{\bibinfo{journal}{npj Quantum Information}}
  \textbf{\bibinfo{volume}{9}}, \bibinfo{pages}{20} (\bibinfo{year}{2023}).

\bibitem{Fickler13642}
\bibinfo{author}{Fickler, R.}, \bibinfo{author}{Campbell, G.},
  \bibinfo{author}{Buchler, B.}, \bibinfo{author}{Lam, P.~K.} \&
  \bibinfo{author}{Zeilinger, A.}
\newblock \bibinfo{title}{Quantum entanglement of angular momentum states with
  quantum numbers up to 10,010}.
\newblock \emph{\bibinfo{journal}{Proceedings of the National Academy of
  Sciences}} \textbf{\bibinfo{volume}{113}}, \bibinfo{pages}{13642--13647}
  (\bibinfo{year}{2016}).

\bibitem{barnett2006resolution}
\bibinfo{author}{Barnett, S.~M.} \& \bibinfo{author}{Zambrini, R.}
\newblock \bibinfo{title}{Resolution in rotation measurements}.
\newblock \emph{\bibinfo{journal}{Journal of Modern Optics}}
  \textbf{\bibinfo{volume}{53}}, \bibinfo{pages}{613--625}
  (\bibinfo{year}{2006}).

\bibitem{jha2011supersensitive}
\bibinfo{author}{Jha, A.~K.}, \bibinfo{author}{Agarwal, G.~S.} \&
  \bibinfo{author}{Boyd, R.~W.}
\newblock \bibinfo{title}{Supersensitive measurement of angular displacements
  using entangled photons}.
\newblock \emph{\bibinfo{journal}{Physical Review A}}
  \textbf{\bibinfo{volume}{83}}, \bibinfo{pages}{053829}
  (\bibinfo{year}{2011}).

\bibitem{PhysRevLett.127.263601}
\bibinfo{author}{Hiekkam\"aki, M.}, \bibinfo{author}{Bouchard, F.} \&
  \bibinfo{author}{Fickler, R.}
\newblock \bibinfo{title}{Photonic angular superresolution using twisted n00n
  states}.
\newblock \emph{\bibinfo{journal}{Phys. Rev. Lett.}}
  \textbf{\bibinfo{volume}{127}}, \bibinfo{pages}{263601}
  (\bibinfo{year}{2021}).
\newblock
  \urlprefix\url{https://link.aps.org/doi/10.1103/PhysRevLett.127.263601}.

\bibitem{helstrom1976quantum}
\bibinfo{author}{Helstrom, C.~W.} \& \bibinfo{author}{Helstrom, C.~W.}
\newblock \emph{\bibinfo{title}{Quantum detection and estimation theory}},
  vol.~\bibinfo{volume}{84} (\bibinfo{publisher}{Academic press New York},
  \bibinfo{year}{1976}).

\bibitem{box2011bayesian}
\bibinfo{author}{Box, G.~E.} \& \bibinfo{author}{Tiao, G.~C.}
\newblock \emph{\bibinfo{title}{Bayesian inference in statistical analysis}},
  vol.~\bibinfo{volume}{40} (\bibinfo{publisher}{John Wiley \& Sons},
  \bibinfo{year}{2011}).

\bibitem{d2022experimental}
\bibinfo{author}{D'Aurelio, S.~E.} \emph{et~al.}
\newblock \bibinfo{title}{Experimental investigation of bayesian bounds in
  multiparameter estimation}.
\newblock \emph{\bibinfo{journal}{Quantum Science and Technology}}
  (\bibinfo{year}{2022}).

\bibitem{granade2012robust}
\bibinfo{author}{Granade, C.~E.}, \bibinfo{author}{Ferrie, C.},
  \bibinfo{author}{Wiebe, N.} \& \bibinfo{author}{Cory, D.~G.}
\newblock \bibinfo{title}{Robust online hamiltonian learning}.
\newblock \emph{\bibinfo{journal}{New Journal of Physics}}
  \textbf{\bibinfo{volume}{14}}, \bibinfo{pages}{103013}
  (\bibinfo{year}{2012}).

\bibitem{marrucci-2006spin-to-orbital}
\bibinfo{author}{Marrucci, L.}, \bibinfo{author}{Manzo, C.} \&
  \bibinfo{author}{Paparo, D.}
\newblock \bibinfo{title}{Optical spin-to-orbital angular momentum conversion
  in inhomogeneous anisotropic media.}
\newblock \emph{\bibinfo{journal}{Phys. Rev. Lett.}}
  \textbf{\bibinfo{volume}{96}}, \bibinfo{pages}{163905}
  (\bibinfo{year}{2006}).

\bibitem{PhysRevLett.124.030501}
\bibinfo{author}{G\'orecki, W.},
  \bibinfo{author}{Demkowicz-Dobrza\ifmmode~\acute{n}\else \'{n}\fi{}ski, R.},
  \bibinfo{author}{Wiseman, H.~M.} \& \bibinfo{author}{Berry, D.~W.}
\newblock \bibinfo{title}{$\ensuremath{\pi}$-corrected heisenberg limit}.
\newblock \emph{\bibinfo{journal}{Phys. Rev. Lett.}}
  \textbf{\bibinfo{volume}{124}}, \bibinfo{pages}{030501}
  (\bibinfo{year}{2020}).
\newblock
  \urlprefix\url{https://link.aps.org/doi/10.1103/PhysRevLett.124.030501}.

\bibitem{krenn2015twisted}
\bibinfo{author}{Krenn, M.}, \bibinfo{author}{Handsteiner, J.},
  \bibinfo{author}{Fink, M.}, \bibinfo{author}{Fickler, R.} \&
  \bibinfo{author}{Zeilinger, A.}
\newblock \bibinfo{title}{Twisted photon entanglement through turbulent air
  across vienna}.
\newblock \emph{\bibinfo{journal}{Proceedings of the National Academy of
  Sciences}} \textbf{\bibinfo{volume}{112}}, \bibinfo{pages}{14197--14201}
  (\bibinfo{year}{2015}).

\bibitem{brady2023entanglement}
\bibinfo{author}{Brady, A.~J.} \emph{et~al.}
\newblock \bibinfo{title}{Entanglement-enhanced optomechanical sensor array
  with application to dark matter searches}.
\newblock \emph{\bibinfo{journal}{Communications Physics}}
  \textbf{\bibinfo{volume}{6}}, \bibinfo{pages}{237} (\bibinfo{year}{2023}).

\bibitem{github}
\bibinfo{author}{Belliardo, F.}
\newblock \bibinfo{title}{Bayesian multiparameter}.
\newblock
  \bibinfo{howpublished}{\url{https://github.com/fedebell/AppuntiDottorato/tree/main/bayesian_multiparameter}}
  (\bibinfo{year}{2022}).

\end{thebibliography}
\end{document}